\documentclass{article}

\usepackage{microtype}
\usepackage{graphicx}
\usepackage{subcaption}
\usepackage{booktabs} 

\usepackage{hyperref}
\usepackage{enumitem}
\usepackage{multirow}
\usepackage[table]{xcolor}

\usepackage{amsmath}
\usepackage{amssymb}
\usepackage{mathtools}
\usepackage{amsthm}

\usepackage[preprint]{icml2026}

\usepackage[capitalize,noabbrev]{cleveref}

\usepackage[textsize=tiny]{todonotes}

\icmltitlerunning{Submission and Formatting Instructions for ICML 2026}

\begin{document}

\twocolumn[
  \icmltitle{Learning to Recommend Multi-Agent Subgraphs from Calling Trees}



  \icmlsetsymbol{equal}{*}

  \begin{icmlauthorlist}
    \icmlauthor{Xinyuan Song}{yyy}
    \icmlauthor{Liang Zhao}{yyy}
  \end{icmlauthorlist}

  \icmlaffiliation{yyy}{Department of Computer Science, Emory University, USA}
  \icmlcorrespondingauthor{Liang Zhao}{liang.zhao@emory.edu}

  \icmlkeywords{Machine Learning, ICML}

  \vskip 0.3in
]



\printAffiliationsAndNotice{}  

\begin{abstract}
Multi-agent systems (MAS) increasingly solve complex tasks by orchestrating agents and tools selected from rapidly growing marketplaces. As these marketplaces expand, many candidates become functionally overlapping, making selection not just a retrieval problem: beyond filtering relevant agents, an orchestrator must choose options that are reliable, compatible with the current execution context, and able to cooperate with other selected agents. Existing recommender systems—largely built for item-level ranking from flat user–item logs—do not directly address the structured, sequential, and interaction-dependent nature of agent orchestration. We address this gap by \textbf{formulating agent recommendation in MAS as a constrained decision problem} and introducing a generic \textbf{constrained recommendation framework}:  which first uses retrieval to build a compact candidate set conditioned on the current subtask and context, and then perform \textbf{utility optimization}, which recommends within this feasible set using a learned scorer that accounts for relevance, reliability, and interaction effects. We ground both the formulation and learning signals in \textbf{historical calling trees}, which capture the execution structure of MAS (parent–child calls, branching dependencies, and local cooperation patterns) beyond what flat logs provide. The framework supports two complementary settings: \textbf{agent-level recommendation} (select the next agent/tool) and \textbf{system-level recommendation} (select a small, connected agent team/subgraph for coordinated execution). To enable systematic evaluation, we construct a unified calling-tree benchmark by normalizing invocation logs from eight heterogeneous multi-agent corpora into a shared structured representation. Experiments show our approach consistently recommends higher-quality agents and more coherent agent systems than strong baseline strategies, improving stability, coordination, and end-to-end execution quality in large-scale orchestration. Code and datasets are available at \url{https://github.com/Hik289/Agent_REC.git}.
\end{abstract}

\section{Introduction}

\begin{figure}[!ht]
\centering
\includegraphics[width=\linewidth]{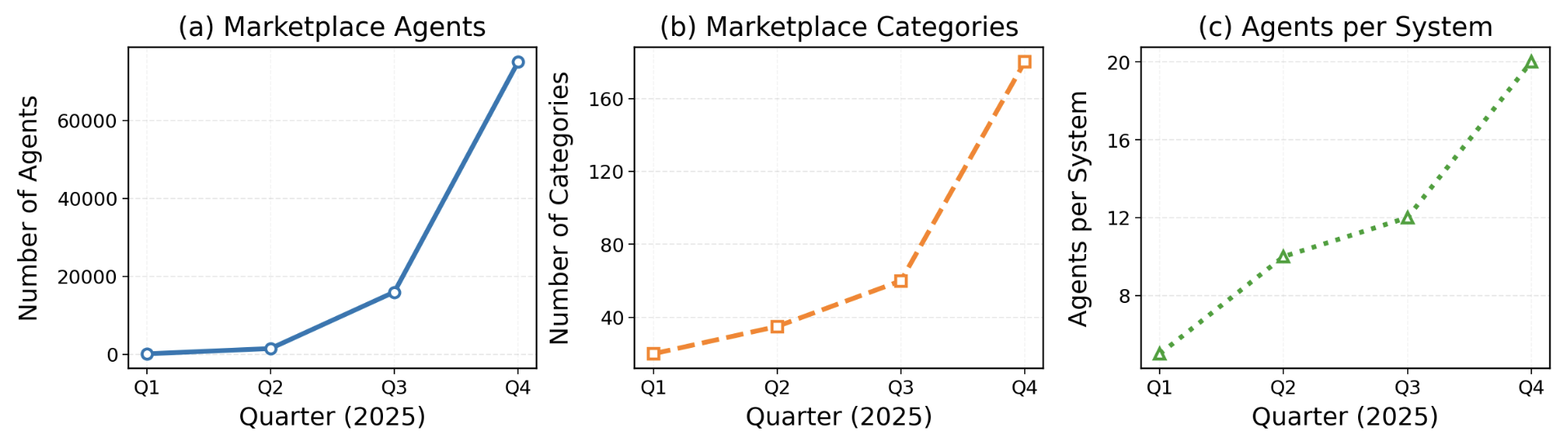}
\caption{Agent marketplace and agent-system growth trends in 2025.
(a) Number of marketplace agents per quarter.
(b) Number of marketplace categories per quarter.
(c) Average number of agents per system per quarter.
All panels report quarterly snapshots in 2025 (Q1--Q4).}
\label{fig:agent_market_growth}
\end{figure}

Multi-agent systems (MAS) have emerged as a practical paradigm for solving complex tasks by orchestrating multiple autonomous agents with diverse capabilities and interfaces~\cite{bansal2025magentic,qin2024toollm}. At their core, such systems operate by selecting and composing agents and tools from a candidate pool to execute a task, with system capability fundamentally bounded by the available agents and tools that can be invoked~\cite{zhou2024agentsurvey}. As agentic systems are increasingly deployed in realistic applications, both the scale of agent systems and the size and diversity of agent marketplaces have grown rapidly, giving rise to pools containing thousands to millions of heterogeneous agents (Figures~\ref{fig:agent_market_growth}). This growth reflects the rising demand for specialized and powerful agents to support increasingly complex agentic workflows, a trend that has become particularly pronounced in recent years and is expected to continue. In such open and evolving ecosystems—where agents are developed independently, updated asynchronously, and subject to practical constraints—multi-agent system performance is increasingly determined by how effectively agents and tools are selected and composed, rather than by the optimization of any single agent in isolation~\cite{liu2024agentmarket,chen2024llmcoordination,park2024graphagent,huang2025callgraph}. As a result, efficiently identifying compatible and effective agents from large-scale marketplaces is becoming a central and nontrivial challenge for modern MAS.

As the number of marketplace agents and tools grows, selecting them cannot be treated as a pure retrieval problem. In practice, an orchestrator will indeed start with retrieval to filter out obviously irrelevant candidates for a requested function, but retrieval only answers which agents/tools are plausible—not which should be chosen. This gap widens as many agents increasingly overlap in functionality: multiple candidates may appear equally relevant by description or embeddings, yet differ substantially in reliability, latency/cost, interface quirks, safety constraints, and how well they cooperate with the rest of the agentic system. Effective selection therefore must go beyond retrieval into a \textbf{recommender-style decision} that conditions on (i) the current system context and goals (task difficulty, budget, time, risk tolerance, downstream dependencies), (ii) social signals such as ratings, reviews, and provider reputation, and (iii) empirical usage experience—e.g., execution logs and calling traces that reveal which agents work well under which contexts and which compositions systematically succeed or fail~\cite{chen2024tracebench,kim2024agenttraces}. These signals enable learning to rank and compose candidates under practical constraints, turning “find relevant agents” into “recommend a reliable, compatible set of agents/tools for this system,” often at the level of agent teams rather than isolated components~\cite{li2024mcpbench,zhang2025graphtracer}.

To close this gap, we formulate \textbf{agent recommendation in multi-agent systems as a constrained decision problem} and propose a generic \textbf{two-stage framework} that makes the core logic explicit: \textbf{first decide what is feasible, then choose what is best}. The motivation is straightforward—when agent pools become large and overlapping in functionality, naive “pick the top-ranked agent” strategies often fail because good agents can still be \textbf{incompatible}, \textbf{unreliable in a given context}, or \textbf{hard to coordinate} with the rest of the system~\cite{zhou2024agentsurvey,li2024mcpbench,shi2025multiagentbench}.
Concretely, at each execution step, after we build a feasible candidate set via retrieval, we then \textbf{recommend within that set} using a learned scoring rule. To make supervision match the reality of orchestration, we ground the formulation in \textbf{historical calling trees} by the system orchestrator, which capture “who calls whom” and the local cooperation patterns that flat logs miss~\cite{chen2024tracebench,hu2024flame,huang2025callgraph}. To cover both common usage modes, we provide two mathematical instantiations of the same framework: \textbf{agent-level recommendation} (choose a single agent/tool next) in the spirit of retrieval-based routing~\cite{qin2024toollm,patil2023gorilla,wu2023autogen}, and \textbf{system-level recommendation} (choose a small connected agent team/subgraph) to reflect that effective coordination is often a property of a group rather than an isolated agent~\cite{kim2024agenttraces,park2024graphagent}. Finally, we build a unified calling-tree benchmark by normalizing logs from eight multi-agent corpora into a shared structured representation~\cite{li2024mcpbench,zhou2025workflowarchive}, enabling consistent evaluation of both settings; experiments show our approach yields more reliable selections and more coherent agent teams than standard baselines.

Conventional recommender systems in e-commerce are designed for \textbf{item recommendation}: given a user (or session) and a large catalog, they rank items by individual relevance using largely \textbf{i.i.d./conditional-independence assumptions and flat interaction logs} (clicks, views, purchases)~\cite{he2017neuralcf}. These methods are highly effective for top-k ranking and retrieval in settings where the user’s utility can be approximated as an aggregate of independent item scores. However, when we move from product catalogs to agent marketplaces, these assumptions become fragile: in a multi-agent system, selecting an agent is not a terminal decision but an intervention that changes execution, induces dependencies among subsequent choices, and can amplify or suppress the usefulness of other agents~\cite{zhou2024agentsurvey,chen2024llmcoordination}.
Applying recommendation to MAS exposes three core mismatches with the e-commerce setting. \textbf{First, the target is not a single “best” agent but a compatible set of agents/tools whose joint behavior is constrained by an interaction graph (dataflow, calling, or coordination structure), making item-level scoring insufficient}~\cite{huang2025callgraph}. \textbf{Second, agent selection is inherently sequential and state-dependent}: each decision alters the system state, reshapes the feasible space of later collaborations, and introduces non-stationarity that violates standard independence assumptions used in conventional ranking pipelines~\cite{chen2024tracebench,hu2024flame}. \textbf{Third, supervision comes from structured execution signals—calling traces, logs, and success/failure outcomes—with latent tree/graph structure}, rather than flat user–item events; exploiting these signals requires reasoning over compositional interactions rather than treating each selection as an isolated label~\cite{kim2024agenttraces,zhang2025graphtracer}. Together, these mismatches imply that directly porting existing e-commerce recommender techniques—even when applied at a subgraph level—is inadequate, motivating recommendation models that explicitly account for \textbf{structured agent interactions, evolving execution context, and system-level compatibility}.


\footnotetext{Data are derived from archived web snapshots (Internet Archive / Wayback Machine) of three platforms: OpenAI GPTs, AWS Marketplace, and Agent.ai. For each quarter of 2025, we estimate the number of agents, categories, and agents per system by counting the corresponding listings in the archived snapshots.}

Our main contributions are summarized as follows:

\begin{itemize}[left=0em]
    \item \textbf{Agent and Agent-System Recommendation for Marketplaces.}
    We formulate agent and agent-system recommendation as a core problem in agent marketplaces and open multi-agent platforms, showing that practical deployment and user-facing acquisition require recommending coherent and executable agent systems rather than isolated agents.

    \item \textbf{Beyond Traditional Recommenders.}
    We identify fundamental limitations of conventional recommender systems in multi-agent settings, including item-level independence assumptions, flat supervision, and the inability to model execution-induced dependencies, and show why these approaches cannot support agent-system recommendation.

    \item \textbf{Constrained Two-Stage Recommendation Framework.}
    We propose a unified constrained two-stage framework that integrates feasibility construction and utility optimization into a divide-and-conquer design, grounded in historical calling trees and supporting both agent-level and agent-system-level recommendation through a shared parameterized scoring model.

    \item \textbf{Structured Dataset.}
    We construct a unified calling-tree dataset spanning eight heterogeneous multi-agent corpora, normalizing diverse execution logs into a shared calling-graph representation with explicit structural dependencies to support constrained recommendation learning.

    \item \textbf{Empirical Validation.}
    We conduct extensive experiments across agent-level and system-level settings, demonstrating that the proposed framework consistently produces more stable, coherent, and higher-utility multi-agent execution plans than existing baselines.
\end{itemize}

\begin{figure*}[!ht]
    \centering
    \includegraphics[width=0.9\textwidth]{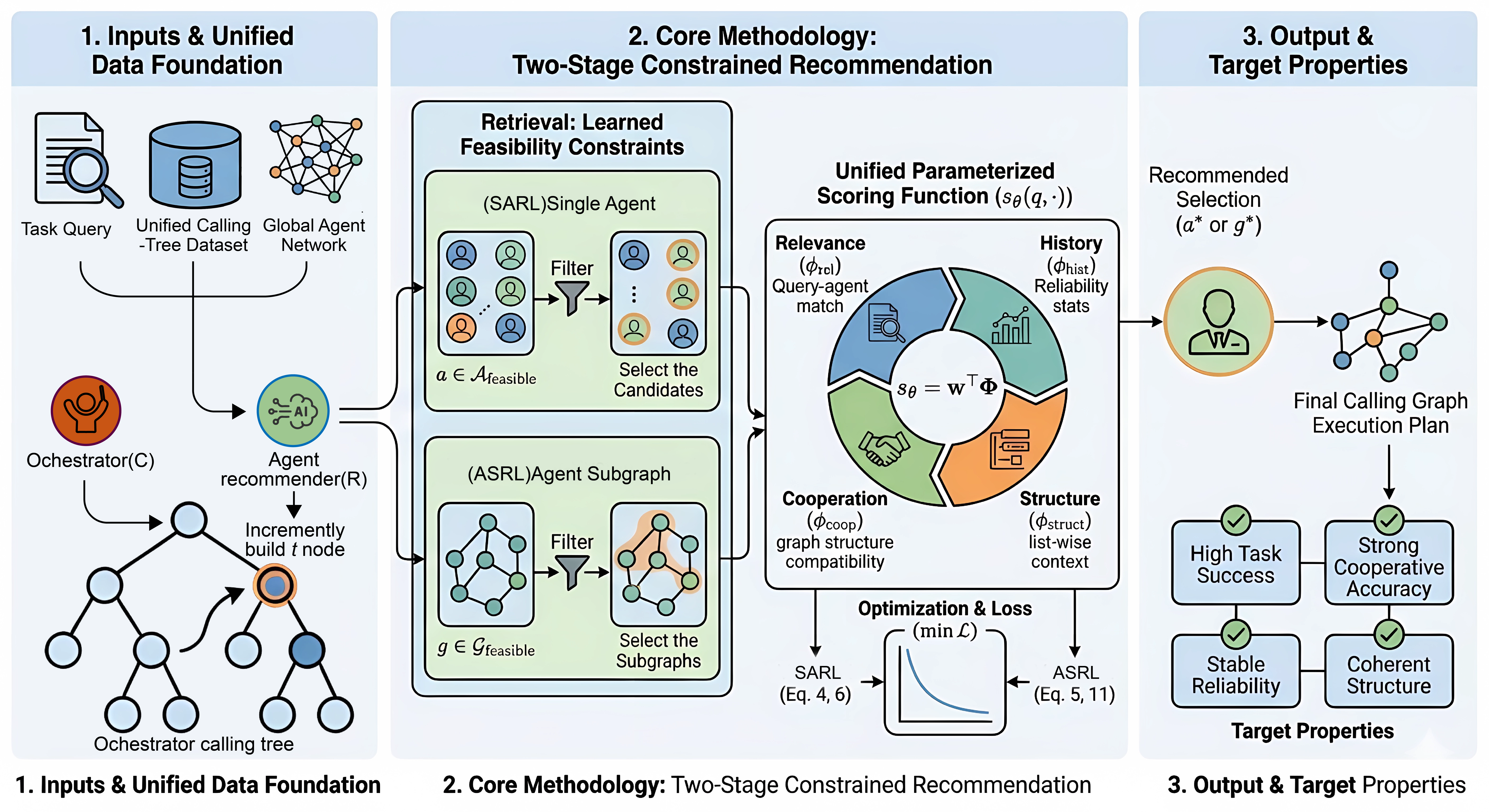}
\caption{
Overview of our two-stage constrained recommendation framework for multi-agent orchestration.
\textbf{(1) Inputs \& data foundation:} a task query, a unified calling-tree dataset, and a global agent network for assisting calling-tree construction.
\textbf{(2) Core methodology (SARL/ASRL):} retrieve feasible agents/subgraphs under learned constraints, then rank with a unified scorer (relevance, history, cooperation, structure).
\textbf{(3) Outputs \& targets:} recommended agents/agent systems and a calling-graph plan, targeting success, reliability, cooperation accuracy, and structural coherence.
}
    \label{fig:method_overview}
    
\end{figure*}

\section{Methodology Overview}

\subsection{Problem Formulation}

We study an \textbf{LLM-based multi-agent recommendation problem} grounded on historical orchestrator execution traces, where agent selection is formulated as a constrained decision problem with learned feasibility sets. We consider a large-scale agent network represented as a graph

\begin{equation}\label{eq:1}
\mathcal{G} = (\mathcal{V}, \mathcal{E}),
\end{equation}
where each node $a_i \in \mathcal{V}$ denotes an agent (or tool), and each edge $(a_i, a_j) \in \mathcal{E}$ encodes a communication, dependency, or interaction relation. Each agent $a_i$ is associated with a base feature representation $\phi(a_i) \in \mathbb{R}^d$ describing its capability, function, or behavioral profile.

A user issues an initial query \(q\). An orchestrator then incrementally decomposes this query into a structured execution plan, formalized as a calling graph
\begin{equation}
\mathcal{T} = (\mathcal{N}, \mathcal{L}),
\end{equation}
where each node \(t \in \mathcal{N}\) represents a concrete intermediate subtask, and edges in \(\mathcal{L}\) encode excutional or logical dependencies between subtasks.

For each node \(t\), the orchestrator maintains a node-specific task description \(q_t\). \(q_t\) encodes the requirements and constraints of the current step in the execution process and is used to determine which agent $a \in \mathcal{V}$, or graph of agents $g \subseteq \mathcal{V}$, should be selected and invoked next. The selection is subject to planning and feasibility constraints informed by historical execution graphs. The definition of \(q_t\) is therefore tied to the position of node \(t\) within \(\mathcal{T}\) and reflects the orchestrator’s current progress in addressing the user query.

\subsection{SARL: Single-Agent Recommendation Learning}\label{version1}

In the first formulation, the orchestrator selects \textbf{individual agents} to populate
the calling graph.
For each node $t$, only a subset of agents is feasible due to execution constraints,
denoted by
\begin{equation}
\mathcal{A}_{\text{feasible}}(t) \subseteq \mathcal{V}.
\end{equation}
Given historical calling traces $\Omega = \{(t, a)\}$ of a fixed agent pool,
where $a$ is the agent invoked at node $t$,
we learn a parametric scoring model that predicts the utility or performance of selecting agent $a$ at state $t$.
The learning objective is defined as

\begin{equation}\label{eq:4}
\begin{aligned}
\min_{\theta} &\sum_{(t,a)\in\Omega}
\ell \left(y_{t,a}, \hat{y}_{t,a}\right)
+ \lambda \|\theta\|^2,\\
\hat{y}_{t,a} &= s_\theta(t, a) \quad\text{s.t. } a \in \mathcal{A}_{\text{feasible}}(t),
\end{aligned}
\end{equation}
where $y_{t,a}$ denotes the observed execution outcome,
and $\ell(\cdot,\cdot)$ is the loss function. \(\lambda \geq 0\) is a regularization coefficient. At inference time, the orchestrator first retrieves a feasible candidate list from $\mathcal{A}_{\text{feasible}}(t)$ and then selects the top-ranked agent to continue the calling graph expansion.

\subsection{ASRL: Agent-System Recommendation Learning}\label{version2}

The second formulation generalizes the decision unit from a single agent
to an \textbf{agent system}, i.e., a coordinated group of agents that jointly
solve a subtask.
We denote such a system by $g \subseteq \mathcal{V}$,
and let $\mathcal{G}_{\text{feasible}}(t)$ be the set of feasible agent systems
at node $t$, each satisfying internal collaboration and execution constraints.

Let $\Omega = \{(t, g)\}$ denote historical traces where an agent system $g$
is invoked at state $t$.
We learn a predictor and optimize
\begin{equation}\label{eq:5}
\begin{aligned}
\min_{\theta} &\sum_{(t,g)\in\Omega}
\ell \left(y_{t,g}, \hat{y}_{t,g}\right)
+ \lambda \|\theta\|^2,\\
\hat{y}_{t,g} &= s_\theta(t, g) \quad \text{s.t. } g \in \mathcal{G}_{\text{feasible}}(t).
\end{aligned}
\end{equation}
Each agent system $g$ is treated as a self-contained collaborative unit
capable of completing the subtask at node $t$.
Compared with agent-level selection, this formulation allows the orchestrator
to directly recommend structured multi-agent solutions,
leading to stronger coordination and improved execution quality.

\section{Method}

\subsection{Single Agent Recommendation learning (SARL)}\label{sec:ac}

To concretely solve the optimization problem formulated in Equation~\eqref{eq:4}, we adopt a two-stage decision architecture that decomposes agent selection into feasibility construction and utility optimization. This design instantiates a divide-and-conquer strategy: rather than optimizing over the full agent pool, we first restrict the action space to a compact, semantically relevant subset, and then perform recommendation within this reduced space.
Specifically, in the retrieval stage, a retrieval module constructs the feasibility constraint $a \in \mathcal{A}_{\text{feasible}}(t)$ by selecting a small candidate set of agents from the full pool. At each decision node $(t, a) \in \Omega$, the orchestrator formulates an end-to-end subtask specification—consisting of a fixed input description and an expected output form—which is represented as a query $q_t$. Standard encoding and retrieval architectures embedding $q_t$ and agent representations into a shared space, from which a top-$K$ candidate list is retrieved. Importantly, this step relies solely on off-the-shelf retrieval mechanisms and does not introduce task-specific retrieval objectives.

Given the resulting feasibility constraint, the recommendation stage then solves Equation~\eqref{eq:4} as a constrained optimization problem over the retrieved agents, selecting the agent that maximizes downstream utility under the current state. This separation allows planning and decision-making to operate over a reduced yet semantically aligned search space, improving efficiency while preserving expressiveness. The detailed formulation and solution of the recommendation stage constitute the focus of the remainder of this section.

\paragraph{SARL Retrieval.}
At each decision node $(t,a)\in\Omega$, the orchestrator selects an agent to execute the current subtask. To this end, the orchestrator defines an end-to-end subtask specification with a fixed input description and an expected output form, represented as a query \(q_t\). Existing encoding and retrieval architectures are then used to embed \(q_t\) and agent representations into a shared space and retrieve a top-\(K\) candidate list from the agent pool, without introducing task-specific retrieval mechanisms. This modular retrieval-based design follows established agent and tool routing systems, including ToolLLM~\cite{qin2024toollm}, Gorilla~\cite{patil2023gorilla}.
From a system perspective, we implement this retrieval-based orchestration process using LangGraph~\cite{chase2022langchain}, which provides a graph-structured execution model aligned with our calling graph abstraction.

\paragraph{SARL Rerank and Recommendation.}
Given the retrieved candidate set $\mathcal{A}_{\text{feasible}}(t)$, the orchestrator selects the final agent by solving a constrained optimization problem restricted to this set. This recommendation stage can incorporate additional criteria such as execution cost, dependency structure, resource availability, and long-horizon planning considerations. We model this decision score using a parameterized scoring function $s_\theta(\cdot,\cdot)$, which is substituted into  Equation~\eqref{eq:4} to define the training objective.

Let $\Omega=\{(t,a^\star)\}$ denote historical orchestrator traces, where $a^\star$ is the agent selected at node $t$. The model parameters $\theta$ are learned by minimizing

\begin{equation}\label{eq:6}
\begin{aligned}
\min_{\theta} &\sum_{(t,a)\in\Omega}
\ell \Big(s_\theta(q_t,a^\star), s_\theta(q_t,a)\Big)
+ \lambda \|\theta\|^2,\\
&\text{s.t. } a \in \mathcal{A}_{\text{feasible}}(t),
\end{aligned}
\end{equation}
where $\ell(\cdot)$ is instantiated as a softmax cross-entropy loss and $\lambda$ is a regularization coefficient. This objective improves the quality of the learned feasibility constraints, while leaving the final constrained optimization and planning decisions to the orchestrator.

Following a single-scorer formulation, we define the agent score as

\begin{equation}
\begin{aligned}
s_\theta(q,a) = \mathrm{Score}_\theta \Big(e_q(q), e_a(a), g_a(a,\mathcal{T}), h_a(a),  c \left(q,\mathcal{L}(q)\right)\Big).
\end{aligned}
\label{eq:single_scorer}
\end{equation}

\begin{itemize}[leftmargin=0em, topsep=0pt, itemsep=2pt, parsep=0pt, labelsep=0em]
  \item $e_q(q)$ encodes the task query.
  \item $e_a(a)$ represents agent capability.
  \item $g_a(a,\mathcal{T})$ captures graph-structured relations in the agent network.
  \item $h_a(a)$ encodes historical reliability statistics.
  \item $c(q,\mathcal{L}(q))$ models list-aware or slate-aware contextual effects induced by the candidate set.
\end{itemize}

\textbf{Special Case:} Equation~\eqref{eq:single_scorer} admits a linear learning-to-rank instantiation by defining an explicit feature vector

\begin{equation}\label{eq:linear_ltr}
\Phi(q,a)=\begin{bmatrix}
\phi_{\text{rel}}(q,a) \\
\phi_{\text{hist}}(a) \\
\phi_{\text{coop}}(a,q,\mathcal{T}) \\
\phi_{\text{struct}}(a,\mathcal{L}(q),\mathcal{T})
\end{bmatrix}.
\end{equation}

and setting $s_\theta(q,a) = \mathbf{w}^\top \Phi(q,a), \quad \theta = \mathbf{w}$.
\begin{itemize}[leftmargin=0em, topsep=0pt, itemsep=2pt, parsep=0pt, labelsep=0em]

  \item $\phi_{\text{rel}}$ corresponds to agent--query relevance.
  \item $\phi_{\text{hist}}$ acts as an item-level reliability prior.
  \item $\phi_{\text{coop}}$ captures graph-aware compatibility.
  \item $\phi_{\text{struct}}$ represents slate-aware or list-wise utility.
\end{itemize}
The parameter vector $\mathbf{w}$ controls the relative contribution of each factor and can be learned using standard learning-to-rank objectives. Under both formulations, retrieval defines the constrained candidate set $\mathcal{A}_{\text{feasible}}(t)$, while agent selection is performed by optimizing a parameterized scoring function within this set. The unified scorer in  Equation~\eqref{eq:single_scorer} provides a general modeling framework, whereas the linear formulation in  Equation~\eqref{eq:linear_ltr} yields a more interpretable and user-friendly special case. The detailed computation of the components $\phi$ is described in Section~\ref{sec:conc}.

\subsection{Agent-system Recommendation Learning (ASRL)}

To extend the above two-stage architecture to the agent-system setting, we apply the same decomposition at the level of \textbf{agent systems}. We consider retrieving a small set of task-relevant agent systems from a large agent network $\mathcal{G}$, where each agent system is a compact connected subgraph that can be executed in a coordinated manner. In \textbf{ASRL} (Section~\ref{version2}), the orchestrator conditions retrieval on a natural-language task description specifying the end-to-end execution requirement, and returns a top-$K$ list of candidate subgraphs. These retrieved agent-system candidates induce a feasibility constraint over executable systems, within which the subsequent system-level recommendation and planning optimize downstream utility.

At node $t$, recall $\mathcal{G}=(\mathcal{V},\mathcal{E})$ denote a large agent graph in Equation~\eqref{eq:1}. Given a task query $q_t$ expressed in natural language, the objective is to retrieve a ranked list of compact agent-system candidates

\begin{equation}
\begin{aligned}
&\mathcal{G}_{\mathrm{feasible}}(t) = \{\mathcal{G}_1, \dots, \mathcal{G}_M\}, \\
&\mathcal{G}_m = (\mathcal{S}_m, \mathcal{E}_m), \qquad
|\mathcal{S}_m| \ll |\mathcal{V}|,
\end{aligned}
\end{equation}
where each subgraph $\mathcal{G}_m$ is semantically aligned with the task query $q_t$ and exhibits sufficient internal connectivity to support coordinated execution.

\paragraph{ASRL Retrieval.}
At each decision node $t$, the orchestrator retrieves a small set of task-relevant \textbf{agent systems} to execute the current subtask. This step follows the same modular retrieval-based paradigm adopted in Section~\ref{sec:ac}, where a natural-language subtask query $q_t$ is used to retrieve a top-$K$ list of compact, semantically aligned agent systems from the agent network. The retrieved agent-system candidates serve as a learned feasibility constraint for subsequent recommendation and planning.

This retrieval paradigm follows graph-based and retrieval-augmented agent frameworks, including GraphRAG~\cite{graphRAG,gretriever}. Retrieval is used to construct a compact and semantically coherent feasible candidate set rather than to directly determine execution. We implement this graph-based orchestration process using LangGraph.

\paragraph{ASRL Rerank and Recommendation.}
Given a retrieved candidate set of agent systems $\mathcal{G}_{\mathrm{feasible}}(t)$, the orchestrator selects the final agent system by solving a constrained optimization problem restricted to this set. Analogous to agent-level recommendation, retrieval scores only reflect semantic relevance, while system-level recommendation can incorporate additional criteria such as execution cost, internal coordination complexity, resource usage, and long-horizon planning considerations across multiple agents. We model this decision using a parameterized scoring function $s_\theta(\cdot,\cdot)$, which is substituted into the system-level  Equation~\eqref{eq:5} to define the training objective.

Let $\Omega=\{(t,g^\star)\}$ denote historical orchestrator traces, where $g^\star$ is the agent system selected at node $t$. The model parameters $\theta$ are learned by minimizing
\begin{equation}\label{eq:graph_ltr_obj}
\begin{aligned}
\min_{\theta}  &
\sum_{(t,g^\star)\in\Omega}
\ell \Big(
s_\theta(q_t,g^\star), s_\theta(q_t,g)
\Big)
+ \lambda \|\theta\|^2,\\
&\text{s.t. } \mathcal{G} \in \mathcal{G}_{\mathrm{feasible}}(t),
\end{aligned}
\end{equation}
where $\ell(\cdot)$ is instantiated as a softmax cross-entropy loss and $\lambda$ is a regularization coefficient. This objective improves the quality of system-level feasibility constraints, while leaving the final constrained optimization and execution to the orchestrator.

Following a single-scorer formulation, we define the system-level score as

\begin{equation}
\begin{aligned}
s_\theta(q,g) &= \mathrm{Score}_\theta \Big(
e_q(q), 
e_g(g), 
g_g(g,\mathcal{T}), \\
&h_g(g), c \left(q,\mathcal{G}_{\mathrm{feasible}}(t)\right)
\Big).
\end{aligned}
\label{eq:single_scorer_graph}
\end{equation}

\begin{itemize}[leftmargin=0em, topsep=0pt, itemsep=2pt, parsep=0pt, labelsep=0em]

  \item $e_q(q)$ encodes the task query and system-level intent provided by the orchestrator.
  \item $e_g(g)$ aggregates agent representations within subgraph $\mathcal{G}$, capturing the collective capability of the agent system.
  \item $g_g(g,\mathcal{T})$ captures internal cooperation patterns and the structural embedding of $g$ within the global graph $\mathcal{T}$.
  \item $h_g(g)$ encodes historical execution statistics and reliability of the agent system.
  \item $c(q,\mathcal{G}_{\mathrm{feasible}}(t))$ models list-aware or slate-aware contextual effects among candidate agent systems.
\end{itemize}

\textbf{Special Case:} Equation~\eqref{eq:single_scorer_graph} admits a linear learning-to-rank instantiation by defining an explicit system-level feature vector

\begin{equation}\label{eq:linear_ltr_graph}
\Phi(q,\mathcal{G})
=
\begin{bmatrix}
\phi_{\text{rel}}(q,g) \\
\phi_{\text{hist}}(g) \\
\phi_{\text{coop}}(g,q,\mathcal{T}) \\
\phi_{\text{struct}}(g,\mathcal{G}_{\mathrm{feasible}}(t),\mathcal{T})
\end{bmatrix},
\end{equation}

with score $s_\theta(q,g) = \mathbf{w}^\top \Phi(q,\mathcal{G}), 
\quad \theta=\mathbf{w}$.
\begin{itemize}[leftmargin=0em, topsep=0pt, itemsep=2pt, parsep=0pt, labelsep=0em]
  \item $\phi_{\text{rel}}(q,g)$ measures semantic alignment between the task query and the agent system.
  \item $\phi_{\text{hist}}(g)$ encodes system reliability and execution history.
  \item $\phi_{\text{coop}}(g,q,\mathcal{T})$ captures internal cooperation and graph-aware compatibility.
  \item $\phi_{\text{struct}}(g,\mathcal{G}_{\mathrm{feasible}}(t),\mathcal{T})$ models list-wise and structural suitability among candidate systems.
\end{itemize}

The parameter vector $\mathbf{w}$ controls the relative contribution of each factor. As in the agent-level case, retrieval defines the constrained candidate set $\mathcal{G}_{\mathrm{feasible}}(t)$, while system selection is performed by optimizing a scoring function within this set. Equation~\eqref{eq:single_scorer_graph} provides a general framework, whereas Equation~\eqref{eq:linear_ltr_graph} yields an linear special case.

Figure~\ref{fig:method_overview} illustrates the overall architecture of our two-stage constrained recommendation framework, highlighting the separation between retrieval-based feasibility construction and constrained agent or agent-system selection.

\subsection{Token Complexity Analysis.}\label{sec:token}

We measure complexity by the number of LLM input tokens consumed at a single decision node $t$.
Let $N$ be the number of items in the agent pool and $L$ be the token length of one item description.
Let the current execution context (e.g., calling graph used as conditioning context) contain $|V_t|$ nodes and let $L_g$ be the average token length of the description attached to one context node. Under a standard linearized prompt construction, the context contributes

\begin{equation}\label{eq:15}
T_{\text{ctx}}(t)=\mathcal{O}\big(|V_t| L_g\big)
\end{equation}
tokens, and evaluating one item against the context costs $T_{\text{item}}(t)=\mathcal{O}\big(L + |V_t| L_g\big)$.

\paragraph{Token complexity of Naive ranking.}
A direct method that scores every item in the pool at node $t$ consumes

\begin{equation}
T_{\text{direct}}(t)=\mathcal{O}\Big(N\big(L + |V_t| L_g\big)\Big).
\end{equation}
\paragraph{Token complexity of our SARL.}
Our method first retrieves a candidate list of size $K$ and then reranks these $K$ items with the same prompt format.
If the retrieval stage scans the full pool using the same item/context representation (or any method whose token cost is upper bounded by that scan),
then the total token cost is

\begin{equation}
\begin{aligned}
T_{\text{2stage}}(t)
=\mathcal{O}\Big((N+K)\big(L + |V_t| L_g\big)\Big),
\end{aligned}
\end{equation}
where the last equality uses $K\le N$. Hence, in big-$\mathcal{O}$ token cost, the two-stage design matches direct retrieval.

\paragraph{Token complexity of our ASRL}
Now each candidate is a subgraph (agent system). Let $M$ be the number of candidate subgraphs in the pool.
Let a candidate subgraph have on average $s$ nodes, and let $L_s$ be the average token length of one node description inside the candidate subgraph.
A linearized representation of one candidate subgraph costs $T_{\text{subgraph}}=\mathcal{O}(s L_s)$,
Based on Equation~\ref{eq:15}, so direct graph retrieval costs

\begin{equation}
T_{\text{direct-graph}}(t)=\mathcal{O}\Big(M\big(s L_s + |V_t| L_g\big)\Big).
\end{equation}
With two-stage retrieval + rerank over $K_g$ candidate subgraphs ($K_g\le M$), the total cost is $T_{\text{2stage-graph}}(t)=\mathcal{O}\Big(M\big(s L_s + |V_t| L_g\big)\Big)$, which is the same big-$\mathcal{O}$ token cost as direct graph retrieval.

\section{Experiments}

We study a new recommendation setting-selecting executable multi-agent subgraphs from historical calling trees-for which we are not aware of established baselines with the same inputs and outputs. Controlled comparisons are therefore conducted between retrieval-only variants and their \textbf{SARL}- and \textbf{ASRL}-augmented counterparts, and selection quality is evaluated using Top-$K$ accuracy under fixed train/test splits. Overall, the problem is formulated and benchmarked at scale, with unified data and evaluation setups provided to support future work.

\subsection{Unified Dataset Construction}

We construct a unified dataset from historical orchestrator traces  $\mathcal{T}$ to support all retrieval and recommendation experiments. Each data instance is derived from past decision nodes and records the task query, the corresponding candidate set induced by planning constraints, and the selected agent or agent system. The detailed construction procedure and per-corpus statistics are provided in Section~\ref{sec:data}. The detailed computation of the components $\phi$ is described in Appendix Section~\ref{sec:conc}.

\subsection{Implementation Details}

Token usage per query scales sharply with the candidate-pool size. Although reranking is LLM-free at inference time, any LLM-based scoring that evaluates many candidates can exceed practical token budgets. This aligns with common coarse-to-fine workflows (shortlist, then rerank) used to keep token cost manageable under large pools. Using Section~\ref{sec:token} and Table~\ref{tab:unified-stats}, this scaling is most severe for \textbf{GUI-360$^\circ$} and \textbf{Seal-Tools}: assuming a conservative 100-token prompt per candidate (tool description plus a short context summary), direct LLM scoring requires $\sim 6.0\times 10^3$ tokens per query on GUI-360$^\circ$ ($60.16\times 100$) and $\sim 1.05\times 10^6$ tokens per query on Seal-Tools ($10{,}453.22\times 100$). Even with shortlist size $K=20$, total cost is still dominated by the initial scan over the full pool, so it continues to scale with candidate-set size. We therefore enforce a fixed token budget per query for fair comparison across architectures, and use GPT-4o for semantic encoding and comparison with standard non-task-specific defaults, top-$p$ disabled, and maximum generation length capped at 128 tokens for auxiliary text generation. Rerank models are trained with a standard ranking loss, learning rate $1\times 10^{-4}$, and batch size 64.

For 6 moderate-scale datasets (Agent-data-protocol~\cite{song2025agentdataprotocolunifying}, Agents\_Failure\_Attribution~\cite{zhang2025whoandwhen}, GTA~\cite{gta2024}, MCPToolBench++~\cite{fan2025mcptoolbenchlargescaleai}, MedAgentBench~\cite{jiang2025medagentbench}, and TRAIL~\cite{deshpande2025trailtracereasoningagentic}), we follow an LLM-style tool-calling protocol and compare three pipelines under the same per-query token budget: (i) \textbf{direct retrieval}, which uses the LLM to score all candidates; (ii) \textbf{coarse-to-fine (ours)}, which first produces a coarse shortlist of $K=20$ candidates and then reranks the shortlist with a second LLM pass using a more detailed prompt; and (iii) a \textbf{cosine-similarity retriever}, which ranks candidates by embedding similarity without invoking the LLM at inference time. For the two large-scale datasets (GUI-360$^\circ$~\cite{mu2024gui360} and Seal-Tools~\cite{wu2024sealtoolsselfinstructtoollearning}), direct LLM retrieval and LLM-based coarse-to-fine reranking violate the fixed token budget implied by Table~\ref{tab:unified-stats}; therefore, we only report results for the cosine-similarity retriever and rerank on these.

\subsection{SARL on 6 moderate datasets}

We study single-agent recommendation under three retrieval/ranking variants: (i) \textbf{direct retrieval}, which uses either embedding similarity or an LLM to recommend candidates against the query; (ii) \textbf{cosine-similarity + Rank}, where Stage~1 retrieves candidates by embedding cosine similarity and Stage~2 re-ranks them using learning to rank model; and (iii) \textbf{LLM + Rank}, which is our \textbf{SARL} method, where Stage~1 retrieval is performed by an LLM-style tool-calling protocol and Stage~2 re-ranking is performed by the same model. We report Top-1 accuracy, and treat Top-1 as the primary metric for final system performance.

\begin{table}[!ht]
\centering

\caption{Top-1 performance (\%) for SARL across datasets. $\uparrow$ indicates that higher is better on standard datasets, while Agents\_Failure\_Attribution is negative where $\downarrow$ is better.}
\label{tab:top1_single}
\resizebox{\linewidth}{!}{
\begin{tabular}{lcccc}
\toprule
\hline
\textbf{Dataset} & \textbf{Cos-sim Direct} &\textbf{Cos-sim+Rank} & \textbf{LLM Direct} & \textbf{SARL (Ours)} \\
\midrule
agent-data\_protocol $\uparrow$ & 40.0 & 34.21 & 70.50 & \textbf{79.85} \\
Agents\_Failure\_Attribution $\downarrow$ & \textcolor{red}{95.3} & 13.82 & 25.31 & \textbf{12.66} \\
GTA $\uparrow$ & 97.2 & 97.13 & \textbf{100.00} & \textbf{100.00} \\
MCPToolBenchPP $\uparrow$ & 66.5 & 45.81 & 76.70 & \textbf{77.57} \\
MedAgentBench $\uparrow$ & 86.4 & 71.32 & \textbf{99.00} & 84.73 \\
trail-benchmark $\uparrow$ & 96.1 & 94.74 & 96.14 & \textbf{98.15} \\
\midrule
\bottomrule
\end{tabular}}

\end{table}

We evaluate single-agent tool selection under three configurations: \textbf{Cos-sim+Rank} (embedding retrieval plus supervised reranking), \textbf{LLM Direct} (one-pass LLM selection), and \textbf{SARL} (LLM-based retrieval followed by reranking). Table~\ref{tab:top1_single} reports Top-1 accuracy. Overall, \textbf{SARL} performs best on agent\_data\_protocol, MCPToolBenchPP, and trail-benchmark, while \textbf{LLM Direct} is strongest on MedAgentBench; both reach 100\% on GTA. Agents\_Failure\_Attribution is a \textbf{negative} dataset where examples are erroneous traces, so lower Top-1 indicates better behavior. In this case, direct cosine-similarity retrieval cannot reliably identify errors and yields an artificially high score (e.g., 95.3\% in \textcolor{red}{red}). Overall, LLM-based selection is more robust than cosine-similarity retrieval on most positive datasets, and reranking provides additional gains when LLM retrieval returns a strong candidate set.

To quantify the recall bottleneck, Table~\ref{tab:retrieval_sr} reports retrieval success rates for cos-similarity based and LLM based retrieval: for example, on Who\&When (Agents\_Failure\_Attribution), a retrieval success rate of 4.01\% reduces \textbf{Cos-similarity + Rank} from 95.25\% (Oracle) to 3.82\%. Agents\_Failure\_Attribution is a negative dataset in which lower values are preferred; the outcome is again dominated by Stage~1 behavior, since retrieving (and thus selecting) the failure-attribution tool more often increases the reported rate. Degradations appear on agent-data\_protocol (40.00\%) and MCPToolBenchPP (66.51\%).
\begin{table}[!ht]
\centering

\caption{\textbf{SARL stage~1 retrieval success rates (SR, \%).}}
\label{tab:retrieval_sr}
\resizebox{0.85\linewidth}{!}{
\begin{tabular}{lcc}
\toprule
\hline
\textbf{Dataset} & \textbf{Embedding Retrieval} & \textbf{LLM Retrieval} \\
\midrule
agent-data\_protocol & 40.00 & 39.92 \\
Agents\_Failure\_Attribution & 3.99 & 3.94 \\
GTA & 97.10 & 97.05 \\
MCPToolBenchPP & 66.47 & 66.49 \\
MedAgentBench & 86.32 & 86.43 \\
trail-benchmark & 100.00 & 100.00 \\
\midrule
\bottomrule
\end{tabular}}

\end{table}

\subsection{ASRL on 6 moderate datasets}

We study multi-agent recommendation at the graph level, where each candidate is a historical tool-calling graph. We evaluate three retrieval/ranking variants: (i) \textbf{direct retrieval}, which uses either graph embeddings Cos-similarity or an LLM Graph retriever to recommend candidate graphs with the query; (ii) \textbf{cosine-similarity + Rank}, where Stage~1 retrieves graphs by embedding cosine similarity and Stage~2 re-ranks them using an learning to rank model; and (iii) \textbf{LLM + Rank}, which is our \textbf{ASRL} method, where Stage~1 retrieval is performed by an LLM-graph retriever and Stage~2 re-ranking is performed by the same model. We report Top-1 accuracy as the primary metric.

\begin{table}[!ht]
\centering

\caption{Multi-agent (graph) Top-1 accuracy (\%).}
\label{tab:graph_top1}
\resizebox{\linewidth}{!}{
\begin{tabular}{lcccc}
\toprule
\hline
\textbf{Dataset} & \textbf{Cos-sim Direct} & \textbf{Cos-sim + Rank} & \textbf{LLM Direct} & \textbf{ASRL (Ours)}  \\
\midrule
agent-data\_protocol & 85.50 & 85.50 & 91.33 & \textbf{100.00} \\
Agents\_Failure\_Attribution & 94.3 & 94.27 & 98.14 & \textbf{100.00}  \\
GTA & 92.5 & 92.46 & 87.07 & \textbf{100.00}  \\
MCPToolBenchPP & \textbf{100.00} & \textbf{100.00} & \textbf{100.00} & \textbf{100.00}  \\
MedAgentBench & \textbf{100.00} & \textbf{100.00} & \textbf{100.00} & \textbf{100.00}\\
trail-benchmark & 96.50 & 96.45 & 98.92 & \textbf{100.00}  \\
\midrule
\bottomrule
\end{tabular}}

\end{table}

We report graph-level results under four settings: \textbf{Cos-sim Direct} and \textbf{LLM Direct}, which directly retrieve/select a single graph using embedding similarity or an LLM, and their two-stage counterparts \textbf{Cos-sim Rank} and \textbf{ASRL (Ours)}, which apply a reranker on top of the Stage~1 candidate set. Table~\ref{tab:graph_top1} reports Top-1 accuracy. Overall, \textbf{ASRL (Ours)} achieves the strongest performance, reaching 100\% Top-1 on all datasets, while single-stage retrieval is slightly lower and dataset-dependent. In addition, \textbf{Cos-sim Direct} and \textbf{Cos-sim + Rank} are nearly identical across datasets, indicating limited gains from reranking when Stage~1 already returns a single candidate.

\begin{table}[!ht]
\centering

\caption{ASRL stage~1 retrieval success rates (\%).}
\label{tab:graph_recall}
\resizebox{0.8\linewidth}{!}{
\begin{tabular}{lcc}
\toprule
\hline
\textbf{Dataset} & \textbf{Cos-sim Retrieval} & \textbf{LLM Retrieval} \\
\midrule
agent-data\_protocol & 92.2 & 91.2 \\
Agents\_Failure\_Attribution & 97.2 & 98.77 \\
GTA & 98.6 & 95.02 \\
MCPToolBenchPP & \textbf{100.00} & \textbf{100.00} \\
MedAgentBench & \textbf{100.00} & \textbf{100.00} \\
trail-benchmark & \textbf{100.00} & 99.9  \\
\midrule
\bottomrule
\end{tabular}}

\end{table}

Table~\ref{tab:graph_recall} further confirms the recall-dominant regime: the Top-1 values closely track SR on each dataset (e.g., on agent-data\_protocol, SR is 92.2\% for embedding retrieval and 91.2\% for LLM retrieval, consistent with the corresponding Top-1 in Table~\ref{tab:graph_top1}). Overall, graph-level retrieval achieves uniformly high SR across datasets (often near 100\%), which largely explains the strong end-to-end performance in ASRL relative to agent system retrieval.

\subsection{Results on two large-scale datasets}

Tables~\ref{tab:tool_selection_gui360_sealtools} and~\ref{tab:graph_selection_gui360_sealtools} compare single-agent tool selection and multi-agent graph selection on GUI-360$^\circ$ and Seal-Tools. On GUI-360$^\circ$, the tool inventory is large, making one-pass direct selection particularly challenging: \textbf{Direct Retrieval} achieves only 18.7\% Top-1 for tools and 17.6\% for graphs, while Stage-1 coverage is substantially higher for graphs than tools (83.5\% vs.\ 61.7\%). In contrast, Seal-Tools contains many entries but is less ambiguous for direct selection, yielding stronger \textbf{Direct Retrieval} (91.3\% for tools and 98.1\% for graphs) with near-perfect Stage-1 coverage (98.4\% and 99.6\%). Across both datasets and both problem settings, the two-stage pipeline is highly reliable: \textbf{LTR (Ours)} reaches 99.9--100.0\% Top-1, showing that when the candidate set is retrieved first (Top-10) and then reranked, the final decision becomes robust even in large search spaces.

\begin{table}[!ht]
\centering

\caption{\textbf{SARL} Retrieval(Stage~1) success rate and Rank(Stage~2) Top-1 accuracy (\%) on the test split.}
\label{tab:tool_selection_gui360_sealtools}
\resizebox{0.8\linewidth}{!}{
\begin{tabular}{lcc}
\toprule
\hline
\textbf{Metric} & \textbf{GUI-360$^\circ$} & \textbf{Seal-Tools} \\
\midrule
Retrieval success (Top-10) & 61.7 & 98.4 \\
Direct Retrieval & 18.7 & 91.3 \\
SARL(Ours) & \textbf{61.9} & \textbf{99.9} \\
\hline
\bottomrule
\end{tabular}}

\end{table}

\begin{table}[!ht]
\centering

\caption{\textbf{ASRL}graph Retrieval(Stage~1) success rate and Rank(Stage~2) Top-1 accuracy (\%) on the test split.}
\label{tab:graph_selection_gui360_sealtools}
\resizebox{0.8\linewidth}{!}{
\begin{tabular}{lcc}
\toprule
\hline
\textbf{Metric} & \textbf{GUI-360$^\circ$} & \textbf{Seal-Tools} \\
\midrule
Retrieval success (Top-10) & 83.5 & 99.6 \\
Direct Retrieval & 17.6 & 98.1 \\
ASRL(Ours) & \textbf{100.0} & \textbf{100.0} \\
\hline
\bottomrule
\end{tabular}}

\end{table}

\subsection{Single-Agent(SARL) vs.\ Multi-Agent(ASRL) Results}

\begin{figure}[!ht]
    \centering
    
    \includegraphics[width=\linewidth]{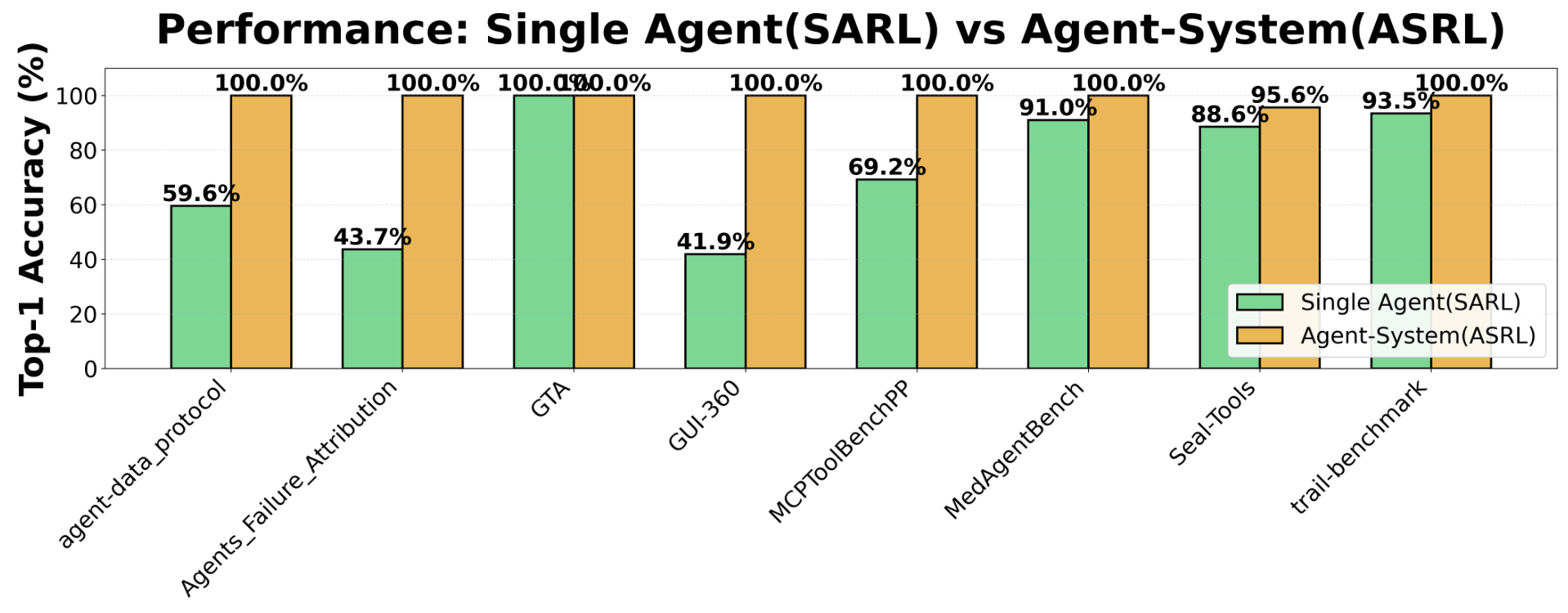}
    
    \caption{Learning-to-rank performance comparison: SARL vs.\ multi-agent ASRL.}
    \label{fig:ltr_bar}
    
\end{figure}

Figure~\ref{fig:ltr_bar} compares SARL and ASRL in terms of Top-1 accuracy. SARL improves the average Top-1 accuracy from \textbf{53.3\%} to \textbf{73.4\%} (about \textbf{+20.2pp}), indicating that supervised ranking can mitigate tool-level ambiguity and partially offset imperfect retrieval. In contrast, ASRL achieves \textbf{near-perfect} Top-1 accuracy across datasets, demonstrating a clear advantage of multi-agent (graph-level) recommendation: richer structured context yields higher candidate recall and reduces semantic ambiguity, allowing the learned ranker to identify the correct graph reliably.

\section{Conclusion}

This work studies agent-team recommendation by unifying heterogeneous multi-agent logs into a graph-based representation and learning scoring objectives that capture semantic relevance, cooperative structure, historical reliability, and call-tree geometry. Building on this representation, we design four complementary scoring components to model alignment, synergy, reliability, and global structural quality. Across diverse benchmarks, our approach consistently outperforms non-cooperative, centralized, factorized, and history-only baselines in task success, relevance, cooperation accuracy, stability, and graph coherence, showing that reliable multi-agent recommendation requires reasoning over full subgraphs and that call-tree supervision provides effective signals for constructing coherent agent teams.

\section{Impact Statement}
This work can improve the reliability and coordination of large-scale multi-agent systems (MAS) by recommending agents/tools (and small agent teams) using constrained decision-making grounded in historical calling trees, which may reduce orchestration failures in practical applications such as automation and decision support; however, it may also amplify biases in past execution logs (e.g., popularity and incumbent advantage), reduce marketplace diversity by repeatedly selecting the same agents, and be misused to scale harmful automated workflows, so deployments should include explicit safety/budget/policy constraints in the feasible set, robustness checks under agent updates and removals, and auditing for exposure and fairness effects.

\bibliography{reference}
\bibliographystyle{icml2026}

\appendix
\clearpage

\section{Related Work}

\paragraph{LLM-based Recommender Systems.}
Large language models have recently been integrated into recommendation models as general reasoning modules.  
LLM4Rec~\cite{LLM4Rec2023} shows that pretrained LLMs can serve as zero-shot recommenders when user histories are formatted as prompts.  
InstructRec~\cite{wang2023instructrec} aligns LLMs with recommendation tasks through instruction tuning, improving controllability and task adherence.  
ReLLa~\cite{lin2023rella} incorporates retrieval-augmented representation into LLM recommenders to reduce hallucination and improve factual grounding.  
Hybrid architectures such as KGRec-LM~\cite{li2023kgreclm} integrate knowledge graphs with LLM reasoning to strengthen structured preference modeling and multi-hop dependency understanding.  
These works treat the LLM as a single high-capacity reasoning module without explicit graph-based interaction or agent collaboration.

\paragraph{Subgraph-based Recommendation.}
A parallel line of research leverages subgraph extraction to improve reasoning over user--item relations.  
SuGeR~\cite{zhang2022sugersubgraphbasedgraphconvolutional} constructs heterogeneous subgraphs centered on a user--bundle pair and applies relational graph propagation to capture fine-grained multi-level interactions.  
SEPTA~\cite{Peng_2024} unifies graph--text alignment, subgraph retrieval, and reasoning into a single model, using contrastive alignment and vector-database retrieval.  
DialogGSR~\cite{park2024generativesubgraphretrievalknowledge} formulates subgraph retrieval as a generative task using structure-aware linearization and graph-constrained decoding.  
The Subgraph Retriever (SR)~\cite{Zhang_2022} expands multi-hop relational paths with a RoBERTa-based scorer and merges them into compact answer-relevant subgraphs through staged pretraining.  
These systems integrate graph structure and textual semantics but operate without agent-level decomposition.

\paragraph{Agent-based Recommendation Systems.}
More recent works use LLM-driven agents to model user behavior and simulate interactive recommendation.  
Agent4Rec~\cite{zhang2024generativeagentsrecommendation} builds generative \textbf{user agents} equipped with factual and emotional memory modules.  
Agents interact with an adaptive environment in a page-by-page loop—viewing, rating, reflecting, and updating preferences.  
This enables modeling of human-like cognitive dynamics such as satisfaction, preference drift, and emotional response, extending recommendation beyond static offline metrics.

\paragraph{Multi-Agent LLM-based Recommendation.}
Multi-agent architectures extend this idea by coordinating heterogeneous LLM agents.  
Rec4Agentverse~\cite{zhang2024prospectpersonalizedrecommendationlarge} frames recommendation as an ecosystem of User, Agent Item, and Agent Recommender agents.  
These agents collaborate through multi-stage pipelines including user--agent interaction, agent--recommender feedback, and inter-agent cooperation.  
Domain-specialized agent items extract semantic preferences, while the recommender orchestrates goal-driven reasoning.  
This paradigm supports interpretable recommendation through multi-agent dialogue and cross-agent knowledge exchange.

\paragraph{Our Position: Recommendation via Calling Graphs and Tool-Calling Trees.}
While prior agent-based systems mainly rely on natural-language conversation among LLM agents, our work focuses on multi-agent recommendation structured by explicit calling graphs and tool-calling trees.  
In this framework, each agent corresponds to an executable module (retriever, ranker, reasoner, or aggregator), and interactions are represented as directed tool calls forming a hierarchical execution graph.  
This structure allows deterministic tracking of intermediate reasoning states, fine-grained decomposition of recommendation tasks, and controllable orchestration across heterogeneous agents.  
By grounding multi-agent recommendation in a formal graph-of-calls representation, our method provides a systematic and interpretable foundation that integrates LLM reasoning, graph structure, and multi-step cooperative decision making.

\section{Dataset}\label{sec:data}
\subsection{Dataset Construction}

To implement our multi-agent recommendation framework, we require data in a consistent graph-based form that reflects how agents interact, collaborate, and trigger one another during execution. Although existing benchmarks were originally developed for diverse purposes—
including tool learning~\cite{wu2024sealtoolsselfinstructtoollearning,fan2025mcptoolbenchlargescaleai}, 
workflow debugging~\cite{deshpande2025trailtracereasoningagentic,zhang2025whoandwhen}, GUI action prediction~\cite{mu2024gui360}, 
multimodal reasoning~\cite{gta2024}, 
and medical API execution~\cite{jiang2025medagentbench}—
they all implicitly contain \textbf{agent calling histories}: sequences of agent or tool invocations linked
through explicit or implicit dependencies.

To support subgraph retrieval and recommendation over these histories, we convert every dataset into a unified graph representation consisting of:

\begin{itemize}
    \item \textbf{Calling Graph (Calling Tree).}  
    A tree graph rooted at the session orchestrator, where each node corresponds to a single invocation (tool call, API call, or agent action). Edges encode parent--child relations derived from span identifiers, caller indices, or reconstructed dependency order. Calling trees are therefore a special case of calling graphs.

    \item \textbf{Agent Pool.}  
    The set of all distinct agents/tools appearing in a session, annotated with metadata such as agent category, tool type, success status, latency, and resource usage. The pool serves as the candidate set for recommendation.
\end{itemize}

\begin{figure*}[!ht]
    \centering
    \includegraphics[width=0.9\linewidth]{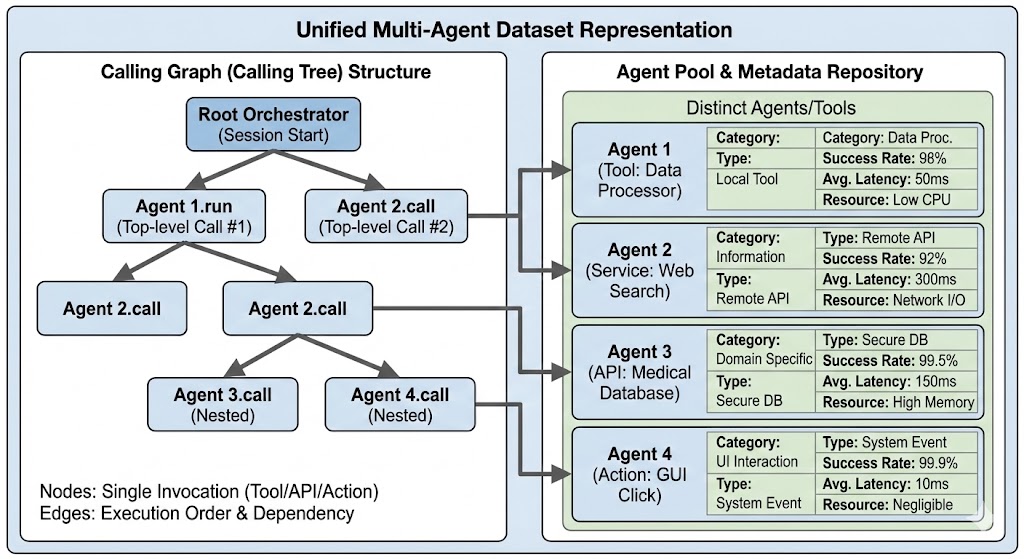}
    \caption{Unified multi-agent dataset representation, including calling-tree structure and agent pool metadata. Nodes represent single tool/API invocations, and edges encode execution ordering and dependency relations.}
    \label{fig:multi_agent_representation}
\end{figure*}

This unified representation enables heterogeneous datasets—
from GUI trajectories~\cite{mu2024gui360},
FHIR-based medical workflows~\cite{jiang2025medagentbench},
multi-step tool-plan corpora~\cite{wu2024sealtoolsselfinstructtoollearning,fan2025mcptoolbenchlargescaleai},
and agent-debug traces~\cite{deshpande2025trailtracereasoningagentic,zhang2025whoandwhen}—
to be analyzed and trained under the same graph-level abstraction.
Figure~\ref{fig:multi_agent_representation} presents the  data structure used throughout the paper to visualize calling graphs.


\subsection{Dataset Details}

The unified calling-graph representation introduced above is constructed from eight datasets that capture tool-use events, multi-step workflows, or multi-agent execution traces. Although these datasets differ widely in domain, purpose, and data format, each provides raw invocation logs from which we extract calling graphs and agent pools after normalization. Below, we summarize the essential characteristics of each dataset after processing, including their scale (\#tools, \#graphs, total nodes) and structural properties (average nodes per graph, average calls per tool) based on the unified statistics.

\paragraph{MCPToolBench++.}
MCPToolBench++~\cite{fan2025mcptoolbenchlargescaleai} is a tool-use benchmark with natural-language tasks paired with tool invocation traces. After processing, it contains 87 tools and 1{,}509 graphs with 1{,}511 total nodes and 1.00 node per graph on average, confirming that it is dominated by single-step tool selection. Tools are invoked 17.37 times on average, indicating moderate reuse across many independent queries.

\paragraph{Seal-Tools.}
Seal-Tools~\cite{wu2024sealtoolsselfinstructtoollearning} is a large-scale, self-instruct tool-learning dataset covering heterogeneous real-world domains and tool schemas. It is the largest tool pool in our benchmark, containing 4{,}076 tools and 14{,}076 graphs with 36{,}099 total nodes. The average graph has 2.56 nodes, reflecting short multi-step workflows, while the average tool is invoked 8.86 times, consistent with a long-tail distribution where many tools are rarely used.

\paragraph{TRAIL (trail-benchmark).}
TRAIL~\cite{deshpande2025trailtracereasoningagentic} is a trace-based agentic reasoning benchmark with multi-step trajectories and tool-use chains. It yields 9 tools and 142 graphs after processing, with 648 total nodes and 4.56 nodes per graph on average, indicating non-trivial multi-step structure. With 72.00 calls per tool on average, TRAIL exhibits heavy reuse of a small tool set.

\paragraph{Who\&When (Agents\_Failure\_Attribution).}
Who\&When~\cite{zhang2025whoandwhen} is a failure-centric dataset that provides incorrect, misleading, or incomplete tool traces for failure attribution and robustness analysis. It corresponds to the \texttt{Agents\_Failure\_Attribution} subset in our unified processing and contains 194 tools and 183 graphs with 1{,}448 total nodes. It has the longest traces among our datasets (7.91 nodes per graph on average), while tools are invoked 7.46 times on average, reflecting diverse tool usage across long failure-oriented execution chains.

\paragraph{AgentBench.}
AgentBench~\cite{liu2023agentbench} is a multi-domain agent evaluation suite spanning OS control, databases, games, web agents, KG reasoning, and household tasks, with environment-driven feedback and multi-step action dependencies. After normalization, we retain its multi-step interaction structure within the unified calling-graph format for downstream retrieval and recommendation experiments.

\paragraph{GUI-360$^\circ$.}
GUI-360$^\circ$~\cite{mu2024gui360} is a large-scale GUI automation dataset consisting of multi-step desktop application interaction trajectories. After processing, it contains 19 tools and 3{,}362 graphs with 22{,}382 total nodes and 6.66 nodes per graph on average, capturing longer realistic GUI workflows. It also has the highest tool reuse (1{,}178.00 calls per tool on average), consistent with a compact GUI action vocabulary repeatedly used across many trajectories.

\paragraph{GTA.}
GTA~\cite{gta2024} is a graph-structured task automation benchmark with multi-step tool-use traces. After processing, it contains 14 tools and 229 graphs with 557 total nodes, yielding 2.43 nodes per graph on average. Tools are invoked 39.79 times on average, reflecting repeated use of a compact perception--logic--operation tool set.

\paragraph{MedAgentBench.}
MedAgentBench~\cite{jiang2025medagentbench} is a medical-domain agent benchmark based on standardized clinical APIs (FHIR), covering multi-step retrieval and decision workflows. After processing, it contains 8 tools and 300 graphs with 660 total nodes and 2.20 nodes per graph on average. Tools are invoked 82.50 times on average, reflecting repeated use of a small set of standardized medical APIs.

\paragraph{ADP (Agent-Data-Protocol).}
ADP~\cite{song2025agentdataprotocolunifying} is a multi-domain trajectory dataset that unifies heterogeneous agent actions into a standardized protocol schema. It corresponds to \texttt{agent-data\_protocol} in our unified processing and contains 83 tools and 84 graphs with 600 total nodes after normalization, with 7.14 nodes per graph on average, indicating relatively long trajectories. Tools are invoked 7.23 times on average, reflecting moderate reuse across diverse agent types and environments.

\subsection{Extraction of Calling Trees and Agent Pools}

We convert each dataset into the unified structure using the following steps:

\begin{enumerate}
    \item \textbf{Normalize raw events.}  
    Each action or tool call is mapped to a canonical event tuple containing timestamp, caller identity, invoked agent, cost, latency, token usage, and metadata.

    \item \textbf{Resolve parent--child links.}
    Parent nodes are determined using (1) \texttt{span\_id} / \texttt{parent\_span\_id} if available; (2) explicit caller indices; or (3) \emph{last-invocation matching}, which assigns each invocation to the most recently preceding invocation in the same session as its parent when explicit linkage is missing.

    \item \textbf{Construct the calling tree.}  
    Events are arranged into a rooted tree, with \texttt{Root} as orchestrator and nodes attached according to resolved parent relationships.

    \item \textbf{Construct the agent pool.}  
    All unique agents/tools invoked in a session form the pool, preserving metadata such as category, success status, and resource usage.

    \item \textbf{Optional pruning.}  
    We remove degenerate trees consisting of a single agent call (e.g., MCPToolBench++ one-step entries) when downstream tasks require structural dependencies.
\end{enumerate}

\subsection{Unified Dataset Statistics}

After converting all datasets into the unified representation of calling trees and agent pools, we compute a consistent set of structural statistics that allows direct comparison across corpora. Table~\ref{tab:unified-stats} reports the number of sessions, total agent invocations, average calling-tree depth, and whether nested calls appear. These statistics quantify the diversity of workflow complexity across the
eight datasets and show how they collectively support multi-step and hierarchical agent reasoning. The final row summarizes the unified dataset obtained by merging all processed corpora, which serves as the training source for our recommendation framework.

\begin{table*}[!ht]
\centering
\caption{Unified calling-tree statistics across all datasets.}
\label{tab:unified-stats}
\resizebox{0.95\textwidth}{!}{
\begin{tabular}{lccccc}
\toprule
\hline
\textbf{Dataset} & \textbf{\#Tools} & \textbf{\#Graphs} & \textbf{\#Nodes} & \textbf{Avg.\ Calls/Tool} & \textbf{Avg.\ Nodes/Graph} \\
\midrule
ADP (Agent-Data-Protocol)~\cite{song2025agentdataprotocolunifying} & 83 & 84 & 600 & 7.23 & 7.14 \\
Who\&When (Agents\_Failure\_Attribution)~\cite{zhang2025whoandwhen} & 194 & 183 & 1{,}448 & 7.46 & 7.91 \\
GTA~\cite{gta2024} & 14 & 229 & 557 & 39.79 & 2.43 \\
GUI-360$^\circ$~\cite{mu2024gui360} & 19 & 3{,}362 & 22{,}382 & 1{,}178.00 & 6.66 \\
MCPToolBench++~\cite{fan2025mcptoolbenchlargescaleai} & 87 & 1{,}509 & 1{,}511 & 17.37 & 1.00 \\
MedAgentBench~\cite{jiang2025medagentbench} & 8 & 300 & 660 & 82.50 & 2.20 \\
Seal-Tools~\cite{wu2024sealtoolsselfinstructtoollearning} & 4{,}076 & 14{,}076 & 36{,}099 & 8.86 & 2.56 \\
TRAIL~\cite{deshpande2025trailtracereasoningagentic} & 9 & 142 & 648 & 72.00 & 4.56 \\
\midrule
\rowcolor{gray!15}
\textbf{Total (All)} & \textbf{4{,}490} & \textbf{19{,}885} & \textbf{63{,}905} & \textbf{14.23} & \textbf{3.21} \\
\hline
\bottomrule
\end{tabular}}
\end{table*}

\subsection{Summary}

Across eight datasets, agent invocation behavior ranges from single-step API calls (MCPToolBench++) to deeply nested multi-tool pipelines (TRAIL, Who\&When, GUI-360$^\circ$). Our unified representation consolidates these heterogeneous logs into consistent calling trees and agent pools, enabling structural modeling, subgraph retrieval, and multi-agent
recommendation on a shared foundation.

\section{Concrete Computation of the Objective Terms}\label{sec:conc}

We formulate both single-agent and multi-agent recommendation as linear learning-to-rank problems, where candidates are scored by a weighted combination of abstract, interpretable feature functions. This formulation follows classical learning-to-rank and information retrieval paradigms, while allowing structural generalization from individual tools to graph-structured execution traces.

\subsection{Unified Scoring Framework}

Given a query $q$ and a candidate $c$ (either a tool or a tool-calling graph), the ranking score is defined as
\begin{equation}
s(q,c) = \mathbf{w}^\top \Phi(q,c),
\end{equation}
where $\mathbf{w}$ is a learned weight vector and $\Phi(q,c)$ is a feature embedding. This linear formulation follows standard practice in learning-to-rank systems due to its interpretability, robustness, and sample efficiency \cite{liu2009learning, burges2005learning}.

We decompose the feature vector into four abstract components:
\begin{equation}
\Phi(q,c) = [\phi_{\text{rel}}, \phi_{\text{hist}}, \phi_{\text{coop}}, \phi_{\text{struct}}],
\end{equation}
each capturing a distinct inductive bias commonly used in retrieval and recommendation systems.

\subsection{Semantic Relevance Feature}

The relevance feature $\phi_{\text{rel}}$ measures semantic alignment between the query and the candidate representation:
\begin{equation}
\phi_{\text{rel}}(q,c) = \mathrm{sim}(\mathcal{E}(q), \mathcal{E}(c)),
\end{equation}
where $\mathcal{E}(\cdot)$ denotes a semantic encoder and $\mathrm{sim}(\cdot,\cdot)$ is a normalized similarity function (cosine similarity in our implementation). Semantic embedding--based relevance is standard in neural information retrieval and dense retrieval systems \cite{karpukhin2020dense, guo2020survey}.

In the single-agent setting, $\mathcal{E}(c)$ encodes tool descriptions. In the multi-agent setting, $\mathcal{E}(c)$ encodes a serialized graph obtained by a root-to-leaf depth-first traversal that lists each node's tool name and natural-language description, together with explicit edge markers (e.g., \texttt{CALLER} $\rightarrow$ \texttt{CALLEE}) to preserve the calling structure, following text-based graph encoding practices \cite{wu2020comprehensive}.

\subsection{Historical Reliability Feature}

The historical reliability feature $\phi_{\text{hist}}$ captures prior performance or trustworthiness of a candidate:
\begin{equation}
\phi_{\text{hist}}(c) = r(c),
\end{equation}
where $r(c) \in [0,1]$ is estimated from historical usage statistics or empirical success rates. Reliability-based priors are widely used in recommender systems and cooperative filtering to stabilize ranking under sparse supervision \cite{koren2009matrix, herlocker2004evaluating}.

\subsection{Cooperation and Compatibility Feature}

The cooperation feature $\phi_{\text{coop}}$ measures query--candidate compatibility beyond semantic relevance:
\begin{equation}
\phi_{\text{coop}}(q,c) = f_{\text{coop}}(q, \mathcal{S}(c)),
\end{equation}
where $\mathcal{S}(c)$ denotes the internal structure of the candidate. In the single-agent case, we instantiate $f_{\text{coop}}$ as simple lexical/semantic overlap signals between the query and the tool name/tags (e.g., keyword overlap and embedding similarity). In the multi-agent case, $\mathcal{S}(c)$ is a vector of graph statistics, and we set
\begin{equation}
f_{\text{coop}}(q,\mathcal{S}(c)) = \mathbf{w}_{\text{coop}}^\top \mathcal{S}(c),
\end{equation}
with $\mathcal{S}(c) = [ |V|,\ |E|,\ \mathrm{depth},\ \mathrm{branch},\ \mathrm{tool\_uniq} ]$ denoting the number of nodes, number of edges, maximum depth from the root, average branching factor, and the number of unique tools in the graph, respectively. Similar structural compatibility signals have been used in graph-based retrieval and recommendation \cite{ying2018graph, hamilton2017representation}.

\subsection{Structural Utility Feature}

The structural feature $\phi_{\text{struct}}$ captures query-independent properties of the candidate:
\begin{equation}
\phi_{\text{struct}}(c) = g(\mathcal{G}(c)),
\end{equation}
where $\mathcal{G}(c)$ denotes the underlying structure. We instantiate $g(\cdot)$ as a bounded linear score over normalized graph statistics:
\begin{equation}
g(\mathcal{G}(c)) = \sigma \left(\mathbf{w}_{\text{struct}}^\top \mathbf{s}(c)\right),
\end{equation}
where $\sigma(\cdot)$ is the logistic function and $\mathbf{s}(c)$ reuses the same normalized statistics as above (and may additionally include a simple density term $2|E|/(|V|(|V|-1))$). This bounded design prevents overly large structural scores while retaining informative topology-level signals \cite{kipf2016semi, wu2020comprehensive}.

\subsection{Discussion}

This abstract formulation unifies single-agent and multi-agent ranking under a common linear framework, while allowing different instantiations of $\mathcal{E}(\cdot)$ and $\mathcal{G}(\cdot)$ depending on candidate type. The key distinction lies not in the ranking model itself, but in the representational richness of the candidate space: graph-structured candidates admit more informative structural and cooperative features, which empirically leads to superior performance in complex and adversarial scenarios.

\end{document}